\documentclass{article}
\usepackage{siunitx}
\usepackage{spconf,amsmath,graphicx}
\usepackage{booktabs}
\usepackage{stfloats}
\usepackage{gensymb}
\usepackage{soul}
\usepackage{url}
\usepackage{hyperref}
\usepackage{amssymb}
\usepackage{pifont}
\usepackage[table]{xcolor}

\hypersetup{
    colorlinks=true,
    linkcolor=blue,
    citecolor=blue, 
}

\usepackage[commentmarkup=uwave]{changes}
\definechangesauthor[name={Santiago Lopez}, color=red]{slt}

\title{Diffusion-Based Limited-Angle CT Reconstruction under Noisy Conditions}
\name{\begin{tabular}{c}
Jiaqi Guo$^{1\dagger}$\thanks{© 20XX IEEE. Personal use of this material is permitted. Permission from IEEE must be obtained for all other uses, in any current or future media, including reprinting/republishing this material for advertising or promotional purposes, creating new collective works, for resale or redistribution to servers or lists, or reuse of any copyrighted component of this work in other works.}
\thanks{$^{\dagger}$ Corresponding author.}, Santiago López-Tapia$^{1}$
\end{tabular}}

\address{$^{1}$ Dept. of Electrical and Computer Engineering, Northwestern University, Evanston, IL, USA}

\begin{document}
%
\maketitle
\begin{abstract}
Limited-Angle Computed Tomography (LACT) is a challenging inverse problem where missing angular projections lead to incomplete sinograms and severe artifacts in the reconstructed images. While recent learning-based methods have demonstrated effectiveness, most of them assume ideal, noise-free measurements and fail to address the impact of measurement noise. To overcome this limitation, we treat LACT as a sinogram inpainting task and propose a diffusion-based framework that completes missing angular views using a Mean-Reverting Stochastic Differential Equation (MR-SDE) formulation. To improve robustness under realistic noise, we propose \texttt{RNSD$^+$}, a novel noise-aware rectification mechanism that explicitly models inference-time uncertainty, enabling reliable and robust reconstruction. Extensive experiments demonstrate that our method consistently surpasses baseline models in data consistency and perceptual quality, and generalizes well across varying noise intensity and acquisition scenarios. 
\end{abstract}
\section{Introduction}
\begin{figure*}[htb]\label{OVERALL_structure}
  \centering
  \includegraphics[width=0.95\textwidth]{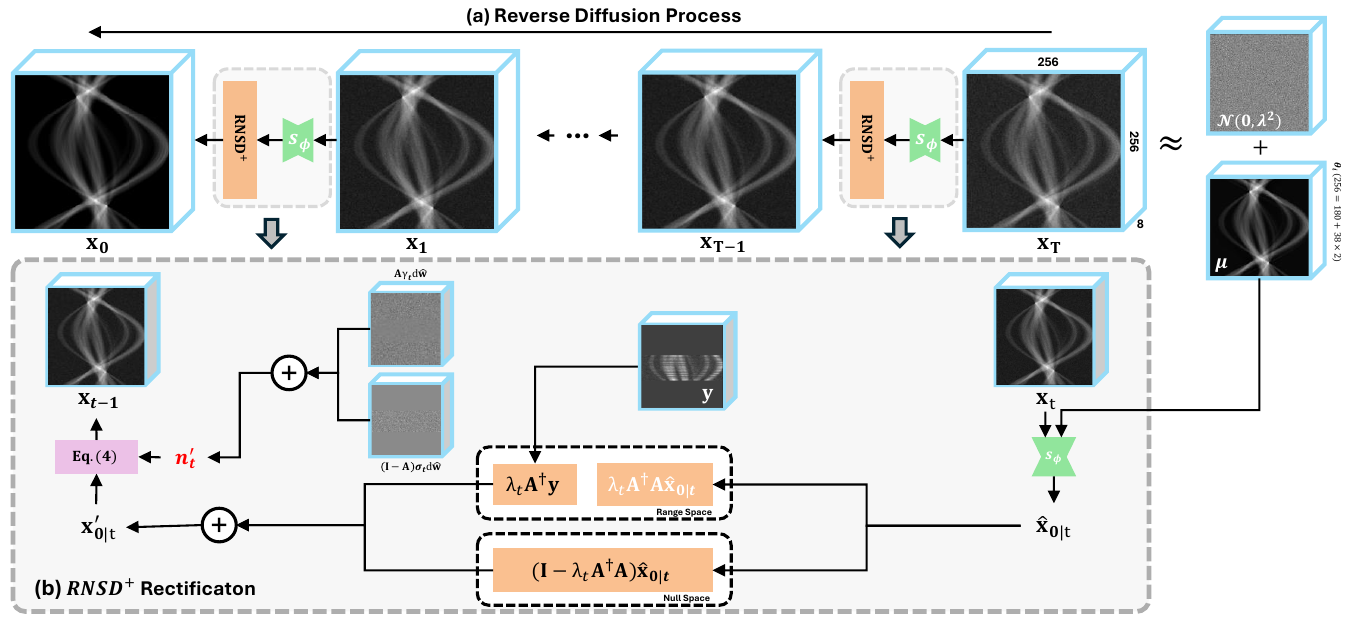}
  \caption{\textbf{Illustration of our proposed LACT reconstruction pipeline.} (a) MR-SDE is employed to restore missing angular information via a mean-reverting diffusion process. The input to the system includes a raw reconstruction containing low-frequency information, obtained using NafNet. This raw reconstruction serves as both the mean $\mu$ in the diffusion process and the conditional input to the network. (b) During inference, we iteratively apply RNSD$^+$ rectification to progressively enforce data consistency.
}
\end{figure*}
\label{sec:intro}
Tomographic reconstruction aims to recover high-dimensional structures from their lower-dimensional projections. This process is governed by the Radon Transform~\cite{radon69u}, which models each projection as a line integral along specific scanning paths. Under ideal full-angle acquisition, these projections sufficiently sample the Fourier space, allowing for accurate image recovery. However, in practical scenarios such as electron tomography, acquisition angles are often restricted, giving rise to the Limited-Angle Computed Tomography (LACT) problem. This inherent incompleteness severely degrades the performance of conventional methods like Filtered Back Projection (FBP)~\cite{dudgeon1983multidimensional}.

In the context of LACT restoration, the goal is to reconstruct a high-fidelity instance $\mathbf{x} \in \mathbb{R}^{M}$ from an incomplete and potentially noisy sinogram $\mathbf{y} \in \mathbb{R}^{N}$. The forward process can be typically formulated as:
\begin{equation}\label{inverse_problem}
    \mathbf{y} = \mathbf{A}\mathbf{x} + \mathbf{n},
\end{equation}
where $\mathbf{A} \in \mathbb{R}^{N \times M}$ denotes the discretized Radon transform operator, and $\mathbf{n}$ represents additive white noise.

Deep learning (DL) has become a powerful tool in image restoration (IR), with early methods predominantly using convolutional neural networks (CNNs) trained under Mean Squared Error (MSE) loss to approximate the inverse of the Radon transform. While effective in terms of computational speed, these approaches often produce overly smooth reconstructions lacking fine structural details. The advent of generative DL models, particularly diffusion models~\cite{ho2020denoising}, has significantly advanced the state of IR. Unlike GANs~\cite{goodfellow2014generative,lin2024drl}, diffusion models capture the full data distribution via a stochastic denoising process, resulting in more accurate and perceptually faithful outputs~\cite{dhariwal2021diffusion, song2020score}.

In the context of Limited-Angle Computed Tomography (LACT), existing learning-based solutions can be broadly categorized into two paradigms. The first line of work focuses on image-domain reconstruction, where the goal is to directly restore the final image from limited-angle projections. For example, Anirudh et al.\cite{anirudh2018lose} propose to bypass sinogram recovery entirely by learning a direct mapping from incomplete sinograms to reconstructed images. More recent approaches such as DOLCE\cite{liu2023dolce} integrate diffusion models with optimization strategies to enhance data fidelity, while RN-SDE~\cite{guo2024rn} leverages Mean-Reverting SDEs and Range-Null Space Decomposition (RNSD) to guide the denoising process. Although effective, these methods typically involve complex operations and are computational expensive due to repeated forward and backward transforms during inference.

The second paradigm reformulates LACT as a sinogram inpainting problem, wherein the missing angular projections are completed directly in the sinogram domain before image reconstruction. Early methods~\cite{ghani2018deep, ghani2019fast, yoo2019sinogram, valat2023sinogram} applied CNNs or GANs to restore corrupted sinograms directly, often with added regularization or continuity constraints. More recently, as an extension of their earlier work~\cite{guo2024rn}, Guo et al.~\cite{guo2025advancinglimitedanglectreconstruction} demonstrated that shifting the restoration process to the sinogram domain simplifies the RNSD-based rectification mechanism and leads to promising results. Despite the growing success of both paradigms, a key limitation persists: most methods assume a noise-free scenario (i.e., $n = 0$), which limits their practical applicability in real-world scenarios where measurements are inherently noisy.

\noindent \textbf{Our Proposed Work:} To develop a more robust and computationally efficient solution to the LACT problem, we extend the framework introduced in~\cite{guo2024rn} and \cite{guo2025advancinglimitedanglectreconstruction} to explicitly account for noise. Specifically, our approach employs the MR-SDE formulation to complete missing angular information through sinogram inpainting. For this purpose, we introduce \texttt{RNSD$^+$}, a noise-aware enhancement of the RNSD rectification mechanism, designed to enforce consistency under non-ideal, noisy conditions. Experimental results show that our improvements lead to significantly better data fidelity over unguided counterparts. We believe this enhancement will broaden the applicability of the method to a wider range of practical imaging scenarios.

\section{Method}
\label{sec:Method}

\subsection{Noisy Sinogram Inpainting with Rectified Diffusion}
\noindent In this work, we leverage the Mean-Reverting Stochastic Differential Equation (MR-SDE) framework proposed by~\cite{luo2023image} to recover incomplete sinograms by modeling the missing angular views. The MR-SDE paradigm captures the forward and reverse evolution of a diffusion process through stochastic differential equations. The forward process incrementally degrades a clean sinogram $\mathbf{x}_\mathbf{0} \sim p(\mathbf{x}_\mathbf{0})$ into a heavily perturbed version $\mathbf{x}_\mathbf{T} \sim p(\mathbf{x}_\mathbf{T})$ over $\mathbf{T}$ steps, according to the following formulation:
\begin{equation}
\mathrm{d}\mathbf{x} = \zeta_t(\mu - \mathbf{x})\,\mathrm{d}t + \sigma_t\,\mathrm{d}\mathbf{w},
\end{equation}
where $\zeta_t$ and $\sigma_t$ are time-varying parameters modulating the rate of mean reversion and noise intensity, respectively, and $\mu$ denotes the terminal mean. The transition distribution at an arbitrary time $t$ given the initial input is:
\begin{equation}
p(\mathbf{x}_t|\mathbf{x}_\mathbf{0})=\mathcal{N}\Big(\mathbf{x}_t\mid \mu+(\mathbf{x}_\mathbf{0}-\mu)e^{-\bar{\zeta}_t},\,\lambda^2\big(1-e^{-2\bar{\zeta}_t}\big)\Big),
\end{equation}
with $\bar{\zeta}_t := \int_{t-1}^{t} \zeta_z\,\mathrm{d}z$ and fixed variance $\lambda^2$. This process is termed \textit{mean-reverting} since $p(\mathbf{x}_t)$ converges to the stationary distribution $\mathcal{N}(\mu, \lambda^2)$ as $t \rightarrow \infty$. In our application, $\mu$ represents the partially observed sinogram, in which the missing views are masked and targeted for recovery.

To reconstruct the full sinogram from the final noisy state $\mathbf{x}_\mathbf{T}$, the reverse SDE is defined as~\cite{luo2023image}:
\begin{equation}\label{reverse_sde}
\mathrm{d}\mathbf{x} = \left[\zeta_t(\mu - \mathbf{x}) - \sigma_t^2 \nabla_{\mathbf{x}} \log p(\mathbf{x}_t)\right]\,\mathrm{d}t + \textcolor{red}{\sigma_t\,\mathrm{d}\widehat{\mathbf{w}}},
\end{equation}
where the score function $\nabla_{\mathbf{x}} \log p(\mathbf{x}_t)$ is approximated by a neural network $\mathbf{s}_\phi(\mathbf{x}, t)$ to maximize the  likelihood of $p(\mathbf{x}_{t-1} \mid \mathbf{x}_t, \mathbf{x}_0)$.

Intuitively, Eq.~\ref{reverse_sde} can also be interpreted as first denoising $\mathbf{x}_t$ to obtain a clean intermediate estimate $\hat{\mathbf{x}}_{0|t}$~\cite{guo2024rn}, followed by injecting a smaller amount of noise, $ \textcolor{red}{\mathbf{n}_t = \sigma_t\mathrm{d}\widehat{\mathbf{w}}}\sim\mathcal{N}(0, \sigma_t^2\mathrm{d}t\mathbf{I})$ to yield the subsequent state $\mathbf{x}_{t-1}$. Based on this process, the RNSD-based rectification scheme~\cite{guo2024rn} in a noisy case can be written as:
\begin{equation}\label{rectification}
    {\mathbf{x}}_{0|t}^{\prime} = \mathbf{A}^{\dagger}(\mathbf{A}\mathbf{x}+\textcolor{blue}{\mathbf{n}}) + (\mathbf{I} - \mathbf{A}^{\dagger}\mathbf{A})\hat{\mathbf{x}}_{0|t},
\end{equation}
where, for sinogram inpainting, ${\mathbf{x}}_{0|t}^{\prime}$ denotes the rectified clean sinogram estimate, and $\mathbf{A}^{\dagger} \in \mathbb{R}^{M \times N}$ is equivalent to the masking operation in sinogram domain. However, this rectification fails to handle noisy observations, as the introduced noise, $\mathbf{A}^{\dagger}\mathbf{n}\in \mathbb{R}^{M}$, can severely disrupt the iterative inference process. Inspired by~\cite{wang2023zero}, we recognize that since $\textcolor{blue}{\mathbf{n}}\sim\mathcal{N}(0, \sigma_y^2\mathbf{I})$ also follows a Gaussian distribution, it can be treated as part of the noise and subsequently removed through diffusion. To this end, we propose to modify the rectification in Eq.~\ref{rectification} and simultaneously adjust the amount of noise later injected ($\textcolor{red}{\mathbf{n}_t\rightarrow\mathbf{n}_t^{\prime}})$, as follows:
\begin{equation}
\begin{aligned}
\mathbf{x}^{\prime}_{0|t} 
&= \hat{\mathbf{x}}_{0|t} - \lambda_t \mathbf{A}^\dagger \left( \mathbf{A} \hat{\mathbf{x}}_{0|t} - \mathbf{y} \right) \\
&= (\mathbf{I} - \lambda_t \mathbf{A}) \hat{\mathbf{x}}_{0|t} - \lambda_t\mathbf{y}, \\
\textcolor{red}{\mathbf{n}_t^{\prime}} &= \left[\mathbf{A}\gamma_t + (\mathbf{I}-\mathbf{A})\sigma_t \right] \mathrm{d}\widehat{\mathbf{w}},
\end{aligned}
\end{equation}
with $\mathrm{d}\widehat{\mathbf{w}} = \sqrt{\mathrm{d}t} \sigma_t \epsilon_t$, where $\epsilon_t \sim \mathcal{N}(\mu, \lambda^2)$, and the remaining components are defined as:
\begin{equation}
\begin{array}{rl}
\lambda_t = &
\begin{cases}
1, & \sigma_t \geq h_t \sigma_y, \\
\beta \sigma_t \sqrt{\mathrm{d}t}/{(h_t \sigma_y)}, & \sigma_t < h_t \sigma_y,
\end{cases} \\[3.5ex]
\multicolumn{2}{c}{
\gamma_t = \sigma_t^2 \mathrm{d}t - (h_t \lambda_t \sigma_y)^2
} \\[1.5ex]
\multicolumn{2}{c}{
h_t = \dfrac{1 -  e^{-2 \zeta_t^{\prime}}}{1 - e^{-2 \bar{\zeta}_t}} e^{-\bar{\zeta}_{t-1}}, \quad 
\zeta_t^{\prime} := \int_{t-1}^t \zeta_z \, \mathrm{d} z
}
\end{array}
\label{eq:full}
\end{equation}
Here, both $\lambda_t$ and $\beta$ scale the range-space correction term $\mathbf{A}^\dagger \left( \mathbf{A} \hat{\mathbf{x}}_{0|t} - \mathbf{y} \right)$. The coefficient $\lambda_t$ is set to ensure that the additional noise introduced by the measurement $\mathbf{y}$ remains below the maximum noise level $\sigma_t$ that the model is trained to denoise, while $\beta\in (0, 1]$ serves as an optional damping factor to enhance the robustness of the rectification process and mitigate errors caused by inaccurate estimation of $\sigma_y$. The parameter $\gamma_t$ accounts for any additional noise needed to align the total noise level with $\sigma_t$; it is set to zero when $\sigma_t \leq h_t \sigma_y$, indicating that no extra noise  is required in this scenario. The term $h_t$ corresponds to the scaling coefficient applied when the estimated clean sample $\hat{\mathbf{x}}_{0|t}$ is inserted into Eq.~\ref{reverse_sde}.

\begin{figure*}[htbp]\label{vis_example}
  \centering
  \includegraphics[width=0.98\textwidth]{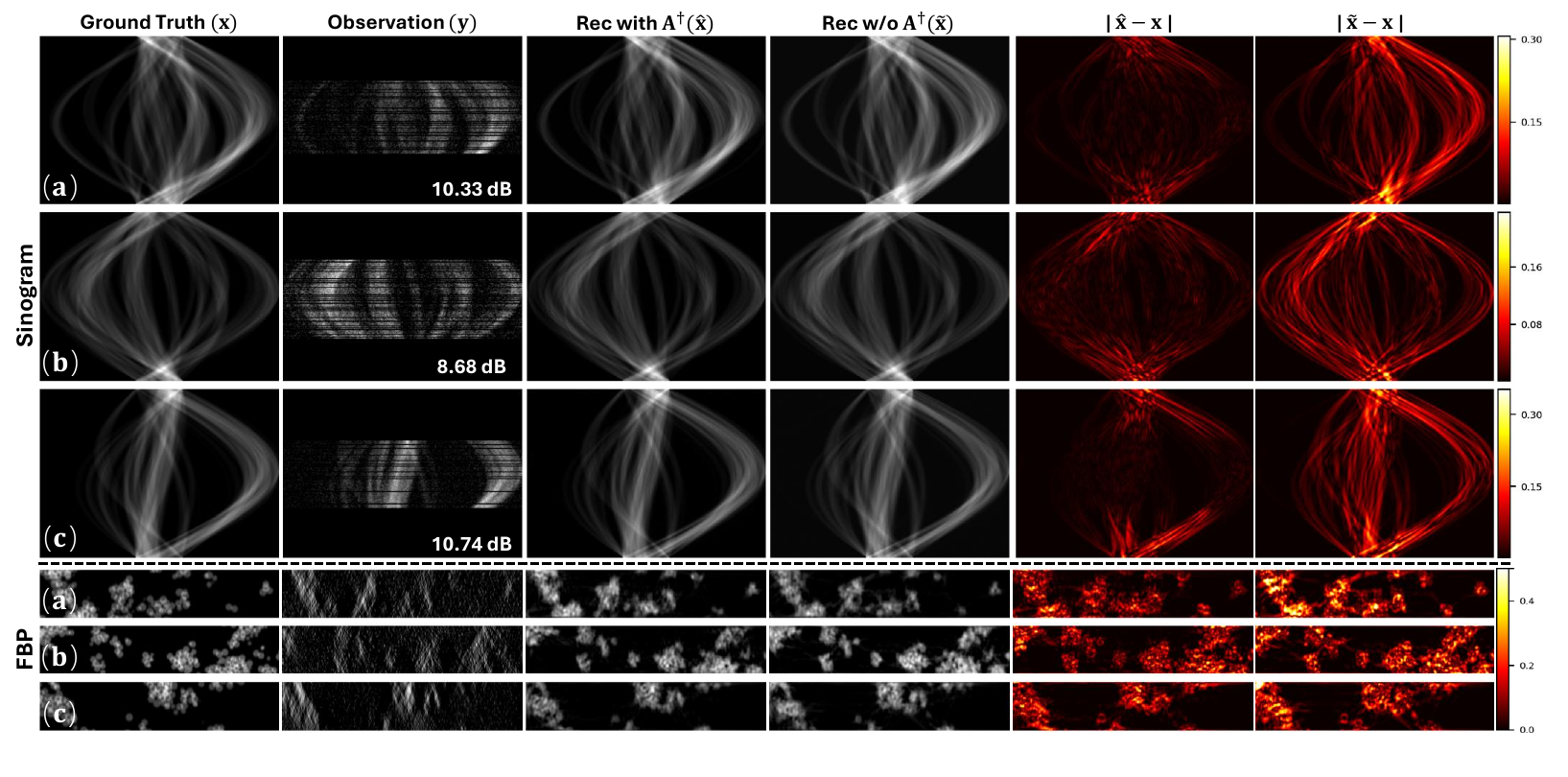}\vspace{-7pt}
    \caption{Illustration of several typical LACT-reconstructed sinograms and their corresponding images obtained via filtered back projection. From left to right: ground truth, noisy observation, reconstruction with \texttt{RNSD$^+$} rectification, reconstruction without rectification (consistency enforced only by MR-SDE), absolute error between reconstruction with \texttt{RNSD$^+$} and ground truth, and absolute error between reconstruction without rectification and ground truth.
    }\vspace{-10pt}
\end{figure*}

\section{Experiment}
\subsection{General Experiment Setting}
\subsubsection{Dataset and Preprocessing}
\noindent\textbf{ChromSTET2025 Dataset:} The \textit{Chromatin Scanning Transmission Electron Tomography (ChromSTET)} dataset is created using the SR-EV~\cite{carignano2024local} model, developed by the Center for Chromatin NanoImaging in Cancer at Northwestern University. This model simulates 3D chromatin structures at nanometer resolution based on known structural and statistical features of chromatin, generating synthetic data that closely resemble real experimental results. The dataset includes $1230$ simulated 3D chromatin volumes from stem cells. Each volume contains $256$ slices, with each slice sized at $50\times256$ pixels, a spatial resolution of 2 nm, and an overall thickness of 100 nm. For training purposes, around $19,680$ sub-volumes with 8 channels are randomly sampled from these structures. To better mimic real ChromSTET imaging conditions, we first generate complete-angle sinograms and then apply a masking process to simulate realistic missing-angle scenarios. Specifically, $61$ scanning angles are randomly selected within the range of $45^\circ$ to $135^\circ$ ($\approx90^\circ$ missing angle), while the rest are discarded. Additionally, Gaussian white noise is added with a signal-to-noise ratio between 5 dB and 15 dB to reflect noise in actual imaging. Finally, $32$ sinograms from different chromatin structures are randomly chosen as the test set, with the rest used for training.

\subsubsection{Basic Experiment Setup}
The MR-SDE model is trained following the configuration in~\cite{guo2024rn}, with the time step parameter fixed at $\mathbf{T}=100$. In addition to the noisy input, a low-fidelity reconstruction generated by NafNet is included as part of the input. This auxiliary input is not required and can be replaced with any other reconstruction method. The model is implemented in PyTorch and trained using four Nvidia Quadro RTX 8000 GPUs in single-precision mode ($\texttt{float32}$). Given the periodicity of sinograms, circular padding is applied along the theta dimension to resize all sinograms to $256\times256$ (\textbf{during training}). The AdamW optimizer is employed with an initial learning rate of $5\times10^{-4}$, adjusted via a cosine annealing scheduler. Training is conducted for 300,000 iterations with a batch size of 16. Unless otherwise specified, the time-travel~\cite{wang2023zero} strategy is adopted by default to mitigate disharmony introduced by rectification.

\subsection{Evaluation and Comparison}
In this section, we compare our proposed method with several limited-angle CT (LACT) reconstruction techniques. Table~\ref{Tab:compare} presents the ten-run average results for PSNR, SSIM, and LPIPS. The evaluated methods include classical approaches (FBP), MMSE-based iterative reconstruction methods (NafNet~\cite{chen2022simple}), and diffusion-based methods (RN-SDE~\cite{guo2025advancinglimitedanglectreconstruction}). It is important to note that most related studies, such as DOLCE~\cite{liu2023dolce}, are conducted under noise-free scenarios. Therefore, they are not included in our comparison. To ensure a fair evaluation across all methods, we use the basic FBP algorithm to back-project each completed sinogram into the image domain.

\begin{table}[htbp]
\caption{\footnotesize Comparison of Different Methods on ChromSTET2025}
\label{Tab:compare}\hspace{-8pt}
\scriptsize
\renewcommand{\arraystretch}{0.8}
\resizebox{0.49\textwidth}{!}{
\setlength{\tabcolsep}{5pt}
\begin{tabular}{lccc}
\toprule
\textbf{Method} & \textbf{PSNR$\uparrow$} & \textbf{SSIM$\uparrow$} & \textbf{LPIPS$\downarrow$} \\
\midrule
\multicolumn{4}{c}{\textit{Sinogram Domain}} \\
\midrule
NafNet~\cite{chen2022simple} & 27.77 & 0.6513 & 0.0445 \\
MRSDE w/o \texttt{RNSD$^+$}~\cite{guo2025advancinglimitedanglectreconstruction} & 32.03 & 0.911 & 0.0300 \\
MRSDE w/ \texttt{RNSD$^+$} & \textbf{35.91} & \textbf{0.950} & \textbf{0.0222} \\
\midrule
\multicolumn{4}{c}{\textit{Image Domain (with FBP)}} \\
\midrule
FBP & 13.24 & 0.191 & 0.5890 \\
NafNet~\cite{chen2022simple} & 19.19 & \textbf{0.680} & 0.1290 \\
MRSDE w/o \texttt{RNSD$^+$}~\cite{guo2025advancinglimitedanglectreconstruction} & 18.75 & 0.578 & 0.1230 \\
MRSDE w/ \texttt{RNSD$^+$} & \textbf{20.43} & 0.665 & \textbf{0.1053} \\
\bottomrule
\end{tabular}
}\vspace{-10pt}
\end{table}

Our method outperforms all baselines in sinogram completion, achieving the best results across metrics. This demonstrates that, when the noise level is known, the proposed RNSD+ correction leads to significant performance gains—approximately 3 dB improvement in PSNR and a 0.04 increase in SSIM. In the image domain, our approach also maintains a clear advantage, yielding notable improvements in both PSNR and LPIPS.

Qualitatively, we provide several typical reconstruction examples using different reconstruction methods, as illustrated in Figure~\ref{vis_example}. It is evident that traditional approaches, such as filtered back projection (FBP, second column), degrade significantly under high noise, resulting in substantial structural loss. In comparison, reconstructions using our proposed \texttt{RNSD$^+$} and those based solely on MR-SDE for enforcing data fidelity (columns 3–6) demonstrate clear differences. Notably, \texttt{RNSD$^+$} considerably improves the alignment between the reconstructed image and the noisy observation, offering significant enhancements over the MR-SDE (w/o rectification) only baseline.

\subsection{Robusteness Analysis}
\begin{figure}[h]
\hspace{-5pt}
  \includegraphics[width=0.49\textwidth]{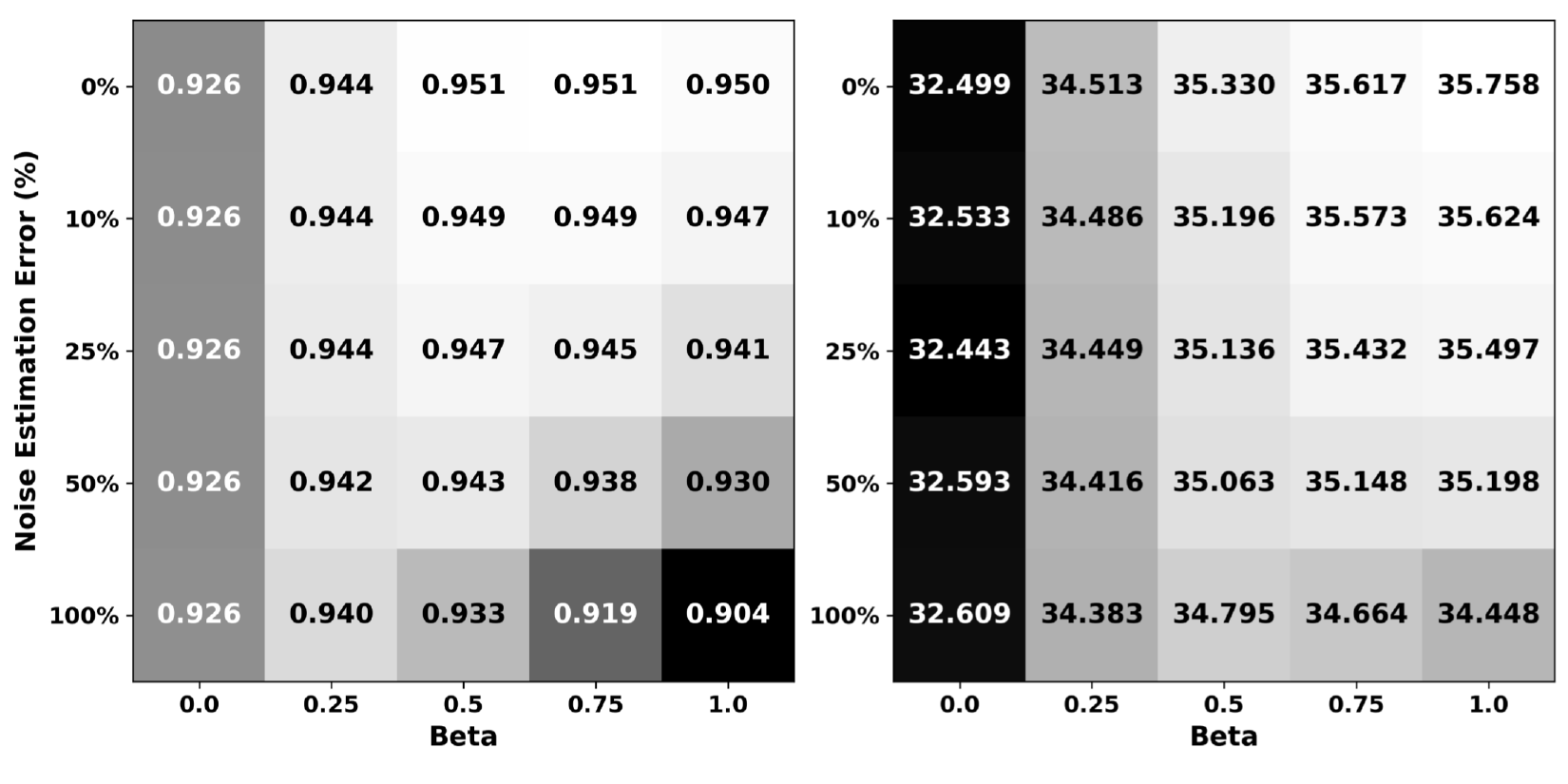}\vspace{-10pt}
    \caption{Evaluation of the robustness of \texttt{RNSD$^+$} under varying levels of noise estimation error (vertical axis: relative error in noise standard deviation) and different choices of $\beta$ (horizontal axis). The left plot shows \textbf{SSIM} results, while the right plot presents \textbf{PSNR} results.}\label{robustness}
\end{figure}
In our experiments, we simulate noisy observations $\mathbf{y}$, allowing for precise estimation of the noise scale within the images. However, in practical scenarios, obtaining an accurate noise estimate is often challenging. To improve the robustness of our method under such uncertainty, we introduce an additional parameter $\beta$ in \texttt{RNSD$^+$} to control the strength of the correction applied from the range-space. Although a smaller $\beta$ may reduce the effectiveness of the rectification ($\beta=0 \rightarrow$ random generation), it provides flexibility in handling noise estimation errors. Here, we evaluate the robustness of our rectification mechanism against noise scale mismatches by injecting artificial noise perturbations, and demonstrate the effectiveness $\beta$.

We present the corresponding evaluation results (averaged over 10 runs) in Figure~\ref{robustness}. As expected, when the noise estimation error increases, setting $\beta=1$, which corresponds to applying the full range-space correction, results in a gradual decline in reconstruction quality. The accumulated noise increasingly disrupts the inference process, leading to the lowest SSIM scores at 100\% error. In addition, we observe a general decreasing trend in both PSNR and SSIM as $\beta$ becomes smaller. When $\beta=0$, no correction is applied, and data consistency is controlled solely by the MR-SDE process. In this setting, the reconstructions are not influenced by the increasing noise in the observations, leading to more random and noise-independent results, as shown in the leftmost column.

\section{Conclusion \& Discussion} \label{sec:foot}
In this paper, we proposed a diffusion-based framework for robust limited-angle computed tomography (LACT) reconstruction under noisy conditions. Our method formulates LACT as a sinogram inpainting task, leveraging a Mean-Reverting Stochastic Differential Equation (MR-SDE) to recover missing angular views by progressively refining sinogram estimates via a reverse diffusion process. To better cope with real-world measurement uncertainty, we introduce an enhanced version of Range-Null Space Decomposition, termed \texttt{RNSD$^+$}, which explicitly incorporates noise modeling and enforces data fidelity throughout inference. Unlike previous approaches that assume ideal conditions ($\mathbf{n}=0$), our framework is designed to handle noisy scenarios with inherent measurement uncertainty, enhancing its practical applicability.

Through extensive experiments on the ChromSTET2025 dataset, we demonstrate that our method consistently outperforms both classical and recent learning-based baselines in terms of structural accuracy and perceptual quality. We show that our rectification strategy maintains robustness against noise scale mismatches, and that the inclusion of a tunable parameter $\beta$ offers a flexible means to control the balance between reconstruction stability and data consistency. These findings underscore the importance of jointly modeling signal and noise during inference, and position our approach as a strong candidate for high-fidelity CT reconstruction in noisy, limited-angle imaging environments.

\bibliographystyle{IEEEbib}

\bibliography{strings,refs}

\end{document}